%% file: PbPbPhotonFFPaper.tex
\pdfoutput=1
\newcommand*{\ATLASLATEXPATH}{}
\documentclass[cernpreprint, texlive=2016, UKenglish]{\ATLASLATEXPATH atlasdoc}

\usepackage{\ATLASLATEXPATH atlaspackage}
\usepackage{\ATLASLATEXPATH atlasbiblatex}

\usepackage{\ATLASLATEXPATH atlasphysics}

\graphicspath{{logos/}{figures/}}

\usepackage{PbPbPhotonFFPaper-defs}

\AtlasTitle{Comparison of fragmentation functions for light-quark- and gluon-dominated jets from $pp$ and Pb+Pb collisions in ATLAS}

\AtlasAbstract{%
Charged-particle fragmentation functions for jets azimuthally balanced
by a high-transverse-momentum, prompt, isolated photon are measured in
$25$~pb$^{-1}$ of $pp$ and $0.49$~nb$^{-1}$ of Pb+Pb collision data at
$5.02$~\TeV\ per nucleon pair recorded with the ATLAS detector at the
Large Hadron Collider. The measurements are compared to predictions of
Monte Carlo generators and to measurements of inclusively selected
jets. In $pp$ collisions, a different jet fragmentation function in
photon-tagged events from that in inclusive jet events arises from the
difference in fragmentation between light quarks and gluons. The
ratios of the fragmentation functions in Pb+Pb events to that in $pp$
events are used to explore the parton color-charge dependence of jet
quenching in the hot medium. In relatively peripheral collisions,
fragmentation functions exhibit a similar modification pattern for
photon-tagged and inclusive jets. However, photon-tagged jets are
observed to have larger modifications than inclusive jets in central
Pb+Pb events.}

\author{The ATLAS Collaboration}

\PreprintIdNumber{CERN-EP-2019-009}

\AtlasJournalRef{Phys.\ Rev.\ Lett. 123, 042001 (2019)}
\AtlasDOI{10.1103/PhysRevLett.123.042001}

\hypersetup{pdftitle={ATLAS document},pdfauthor={The ATLAS Collaboration}}

\begin{document}

\maketitle

Ultrarelativistic nucleus--nucleus collisions create a quark--gluon
plasma, a hot, dense, and long-lived system of deconfined quarks and
gluons. The high density of unscreened color charges causes
hard-scattered partons with large transverse momentum (\pT) to lose
energy as they traverse the medium, a phenomenon referred to as jet
quenching. In lead--lead (Pb+Pb) collisions at the Large Hadron
Collider~(LHC), jet production rates at fixed \pT are suppressed
relative to proton--proton (\pp)
collisions~\cite{HION-2013-06,HION-2011-03,CMS-HIN-13-005,Abelev:2013kqa}. Since
the parton shower develops inside the quark--gluon plasma, the
momentum distributions of hadrons in the quenched jet are also
modified. Measurements of the jet fragmentation function (FF) for
inclusively produced jets in Pb+Pb
collisions~\cite{HION-2015-05,HION-2017-04,CMS-HIN-12-013} exhibit
differences from \pp collisions. In these measurements, jets are
selected by their final-state \pT, i.e.\ after the effects of
quenching, which may result in a bias towards jets that have suffered
only modest modifications and complicates interpretation of the
data~\cite{Casalderrey-Solana:2016jvj,Spousta:2015fca}. Alternatively,
the initial parton \pT can be tagged with a particle unaffected by the
medium, such as a photon
($\gamma$)~\cite{HION-2012-02,HION-2018-06,CMS-HIN-16-002}. The photon
approximately balances the parton \pT before quenching and thus
selects populations of jets in $pp$ and Pb+Pb collisions with
identical initial conditions. A jet recoiling against a prompt photon
is more likely to be initiated by the showering of a light quark,
whereas inclusive jets are mostly initiated by gluons. Thus
$\gamma$-tagged jets can provide information about how energy loss
depends on the color charge of the initiating parton. Finally, the
photon selection equally samples all geometric production points,
whereas the inclusive selection may be biased towards jets which have
lost less energy or were produced near the surface of the
medium~\cite{Agakishiev:2011nb,Renk:2012ve,Milhano:2015mng}.

Many theoretical models of jet quenching have highlighted the value of
$\gamma$-tagged jet
measurements~\cite{Casalderrey-Solana:2015vaa,Kang:2017xnc,KunnawalkamElayavalli:2016ttl},
inviting systematic comparisons of these with inclusive jet
measurements and with theoretical predictions for inclusive and
$\gamma$-tagged jets. The comparisons are best performed if the
measurements are fully corrected for detector effects and presented at
particle level. This Letter presents such a measurement of the FF in
high-\pT jets azimuthally balanced by a prompt, isolated photon in \pp
and Pb+Pb collisions at a center-of-mass energy of $5.02$~\TeV\ per
nucleon pair, using data samples with integrated luminosities of
$25$~pb$^{-1}$ and $0.49$~nb$^{-1}$, respectively. Photon--hadron \pT
correlations in gold--gold collisions were measured at the
Relativistic Heavy Ion Collider~\cite{STAR:2016jdz,Adare:2012qi}. A
measurement of the $\gamma$-tagged jet FF at the LHC compared the FF
at detector level with theoretical calculations that parameterize the
detector smearing effects~\cite{CMS-HIN-16-014}.

Following previous measurements in
ATLAS~\cite{HION-2015-05,HION-2017-04}, the FF for a jet to contain a
charged particle with a given $\pT$, $\eta$ and
$\phi$~\cite{footnote1} is expressed as $D(\pT) =
(1/N_\mathrm{jet})(dN_\mathrm{ch}(\pT)/d\pT)$ or $D(z) =
(1/N_\mathrm{jet})(dN_\mathrm{ch}(z)/dz)$ where $N_\mathrm{jet}$ is
the total number of jets, $N_\mathrm{ch}$ is the number of charged
particles associated with a jet, and the longitudinal momentum
fraction, $z$, is defined as $\pT
\cos\left(\Delta{R}\right) / \pT^\mathrm{jet}$, $\Delta{R} = \smash{ ( (
  \eta^\mathrm{jet} - \eta^\mathrm{part} )^2 + ( \phi^\mathrm{jet}
  - \phi^\mathrm{part} )^2 )^{1/2}}$. Only particles with $\Delta{R} <
  0.4$ are considered.

The principal components of the ATLAS
detector~\cite{PERF-2007-01,Capeans:1291633} used in this measurement
are the inner tracking detector, electromagnetic and hadronic
calorimeters, and an online trigger system. The inner detector is
immersed in a $2$~T axial magnetic field and provides charged-particle
tracking in the range $\left|\eta\right| < 2.5$. It consists of a
high-granularity silicon pixel detector, a silicon microstrip tracker,
and a transition radiation tracker. In the region $\left|\eta\right| <
3.2$, electromagnetic calorimetry is provided by barrel and endcap
high-granularity lead/liquid-argon (LAr) sections divided into three
layers in depth. Hadronic calorimetry is provided by a
steel/scintillator-tile calorimeter, segmented into three barrel
structures within $\left|\eta\right| < 1.7$, and two copper/LAr
hadronic endcap calorimeters, covering the region $1.5
< \left|\eta\right| < 3.2$.  The forward calorimeter is composed of
copper/LAr and tungsten/LAr modules and extends the coverage to
$\left|\eta\right| = 4.9$. During data-taking, events with a high
transverse energy ($E_\mathrm{T}^{\gamma}$) photon are selected using
a two-level trigger system based on energy deposition in the
electromagnetic calorimeter~\cite{TRIG-2016-01}.

Events in Pb+Pb and $pp$ data with photon candidates are selected by
the trigger and are required to contain a vertex reconstructed from
inner-detector tracks. Two centrality classes of Pb+Pb events are
defined using the total transverse energy measured in the forward
calorimeter, \sumETPb. Central events, which are those with a large
nuclear overlap, are defined as those with \sumETPb values in the
highest 30\% percentile (0--30\%) of all Pb+Pb events. Peripheral
events have a \sumETPb value in the $30$--$80$\% percentile and a
smaller nuclear overlap region. The mean number of nucleon--nucleon
collisions in these events is $1080\pm70$ and $135\pm9$, respectively,
evaluated using the Glauber model~\cite{HION-2016-07}.

Monte Carlo (MC) simulations are used to study the performance of the
detector and provide comparisons with data. The main simulation sample
was generated with the \PYTHIA 8.186~\cite{Sjostrand:2007gs}
generator, with the NNPDF23LO parton distribution function (PDF)
set~\cite{Ball:2012cx}, and parameters tuned to reproduce $pp$ data
(``A14'' tune)~\cite{ATL-PHYS-PUB-2014-021}. Events were passed
through a full \textsc{Geant}4 simulation of the
detector~\cite{Agostinelli:2002hh,SOFT-2010-01}, and reconstructed in
the same way as the data. Two million $pp$ events were generated, and
an additional sample of eight million events were overlaid with Pb+Pb
collision data to describe the effects of the underlying event
(UE). Additional samples of \SHERPA 2.1.1~\cite{Gleisberg:2008ta}
events using the CT10 PDF~\cite{Lai:2010vv}
and \HERWIG~7~\cite{Bellm:2015jjp} events with the MMHT H7UE tune and
leading-order PDF set~\cite{Harland-Lang:2014zoa}, which have a
different description of $\gamma$+multijet topologies, quark/gluon jet
composition and hadronization, are used to study systematic
uncertainties. At particle level, jets and photon isolation energies
are defined using stable particles~\cite{footnote2}.


Photons are measured following a procedure used previously in Pb+Pb
collisions~\cite{HION-2012-02,HION-2018-06}, which includes an
event-by-event estimation and subtraction of the UE contribution to
the energy deposited in each calorimeter
cell~\cite{HION-2011-02}. Photon candidates are reconstructed from
clusters of energy in the calorimeter and identified using
requirements on the properties of their showers~\cite{PERF-2017-02}.
Events with a prompt, isolated photon with $E_\mathrm{T}^{\gamma}$ in
the range $80$~\GeV\ to $126$~\GeV\ (chosen to match the range used
in Ref.~\cite{HION-2018-06}) and absolute pseudorapidity smaller than
$2.37$, excluding the region $1.37$--$1.56$ which has more inactive
material, are selected for analysis. The isolation energy,
$E_\mathrm{T}^\mathrm{iso}$, is determined from the sum of the
transverse energy in cells inside a cone size of $\Delta{R}=0.3$
centered on the photon, after subtracting the photon's contribution to
this quantity, and is required to be $E_\mathrm{T}^\mathrm{iso} <
3$~\GeV\ ($<10$~\GeV) in $pp$ (Pb+Pb) collisions.

The combined photon reconstruction and selection efficiencies in $pp$,
peripheral and central Pb+Pb events are $\approx 90$\%, $85$\% and
$65$--$70$\%, and approximately $10\,000$, $1800$ and $6800$ photons
are selected, respectively. The selected sample contains backgrounds
from hadrons and non-isolated photons, called fake photons, that must
be removed statistically. The background contribution is determined
using a double-sideband
approach~\cite{HION-2012-02,STDM-2014-09,STDM-2016-08} in which the
identification and isolation requirements are inverted to select
background-enriched samples. These are used to estimate the purity of
the selection, which is $\approx80$--$94$\% depending on the collision
system.


Jets are measured following the procedure used previously in $pp$ and
Pb+Pb collisions~\cite{HION-2013-06,HION-2013-08,HION-2011-02}. The
anti-$k_{t}$ algorithm~\cite{Cacciari:2011ma} with $R=0.4$ is applied
to $\Delta\eta\times\Delta\phi = 0.1\times0.1$ calorimeter towers.  An
iterative procedure is used to obtain an event-by-event estimate of
the average $\eta$-dependent UE energy density, while excluding jets
from that estimate. The jet kinematics are corrected for this
background and for the detector response using an $\eta$- and
$p_\mathrm{T}$-dependent calibration derived from simulation and
additional small corrections from {\em in situ}
studies~\cite{ATL-PHYS-PUB-2015-015,HION-2017-10}. Jets are required
to have $63~\GeV < \pt^\mathrm{jet} < 144$~\GeV\ and
$\left|\eta^\mathrm{jet}\right| < 2.1$, and be azimuthally balanced
with the photon, with separation $\left|\Delta\phi\right| > 7\pi/8$.
All $\gamma$--jet pairs meeting the criteria are included in the
analysis, but the requirements mainly select topologies with a single
high-\pT balancing jet~\cite{STDM-2015-11,HION-2018-06}. In simulation, the
\pTjet scale is within 1\% of unity, while the
resolution at $\pTjet = 63$~\GeV\ is 21\% in central Pb+Pb events,
12\% in \pp events, and improves with increasing \pTjet. Among these
jets, 73--83\% are quark jets depending on the generator. The jet
flavor is defined by the highest-\pT parton within $\Delta{R} < 0.4$
of the jet~\cite{PERF-2016-04}.

The jet yield $N_\mathrm{jet}$ is corrected for the combinatorial
pairings of the photon with a jet not associated with the
photon-producing hard scattering, and for the contribution of jets
paired with fake photons. The first is evaluated in the data-overlay
simulation and subtracted on a per-photon basis. The second is
subtracted by measuring this yield in the background-dominated
sidebands described above and scaling it to match the determined
impurity. After these background corrections, the yields are corrected
for the effects of bin migration, which are small due to the
large \pTjet range of the measurement relative to the resolution.


The FFs $D(z)$ and $D(\pT)$ are measured using the differential yield
of charged particles with $p_\mathrm{T} > 1$~\GeV, $N_\mathrm{ch}$,
within $\gamma$-balancing jets, divided by the total jet yield
$N_\mathrm{jet}$. This approach was used in previous
measurements~\cite{HION-2015-05,HION-2015-04} and is needed, together
with the unfolding procedure described below, to account for the
simultaneous bin migration in the jet and particle kinematic
variables, which is correlated through the fragmentation of each
jet. Charged-particle tracks are reconstructed from hits in the inner
detector using an algorithm that is optimized for the high-occupancy
conditions in Pb+Pb collisions~\cite{HION-2011-03,HION-2017-04}.  They
are required to meet several criteria including a minimum number of
hits, the presence of hits predicted by the algorithm, and a small
distance-of-closest approach to the vertex.


The raw charged-particle yield $N_\mathrm{ch}(z)$ or
$N_\mathrm{ch}(\pT)$ is initially determined by measuring the
two-dimensional $(\pT^\mathrm{jet}, \pT)$ or $(\pT^\mathrm{jet}, z)$
distribution. Each entry is corrected for the tracking efficiency at
the given \pT and $\eta$, which varies from 60\% to 80\% depending on
occupancy and pseudorapidity. Three background contributions are
estimated and are subtracted statistically: (1) UE particles and
misreconstructed or secondary tracks, estimated using the rate of
tracks not matched to a generated particle in the data-overlay
simulation, (2) charged particles in jets not produced in the same
hard process as the photon, also estimated in simulation, and (3) the
charged-particle yield in jets correlated with fake photons,
determined using the sideband approach described above.

The two-dimensional yield is corrected for bin migration along both
axes using a Bayesian unfolding
procedure~\cite{DAgostini:1994fjx,Adye:2011gm} as in previous dijet
and $\gamma$--jet measurements~\cite{HION-2012-11,HION-2018-06}. The
simulated \pTjet distributions are reweighted to match those in data,
and the number of unfolding iterations is chosen to minimize the
combination of the total statistical uncertainty and residual
sensitivity to the assumed prior distribution. Due to the large size
of the kinematic bins relative to the experimental resolution, the
unfolding changes the yields by typically $5$\% ($10$\%) in $pp$
(Pb+Pb) collisions. This procedure is further validated with a test
performed by dividing the simulated events into statistically
independent halves.

\begin{figure*}[t]
\begin{center}
\includegraphics[width=0.49\textwidth]{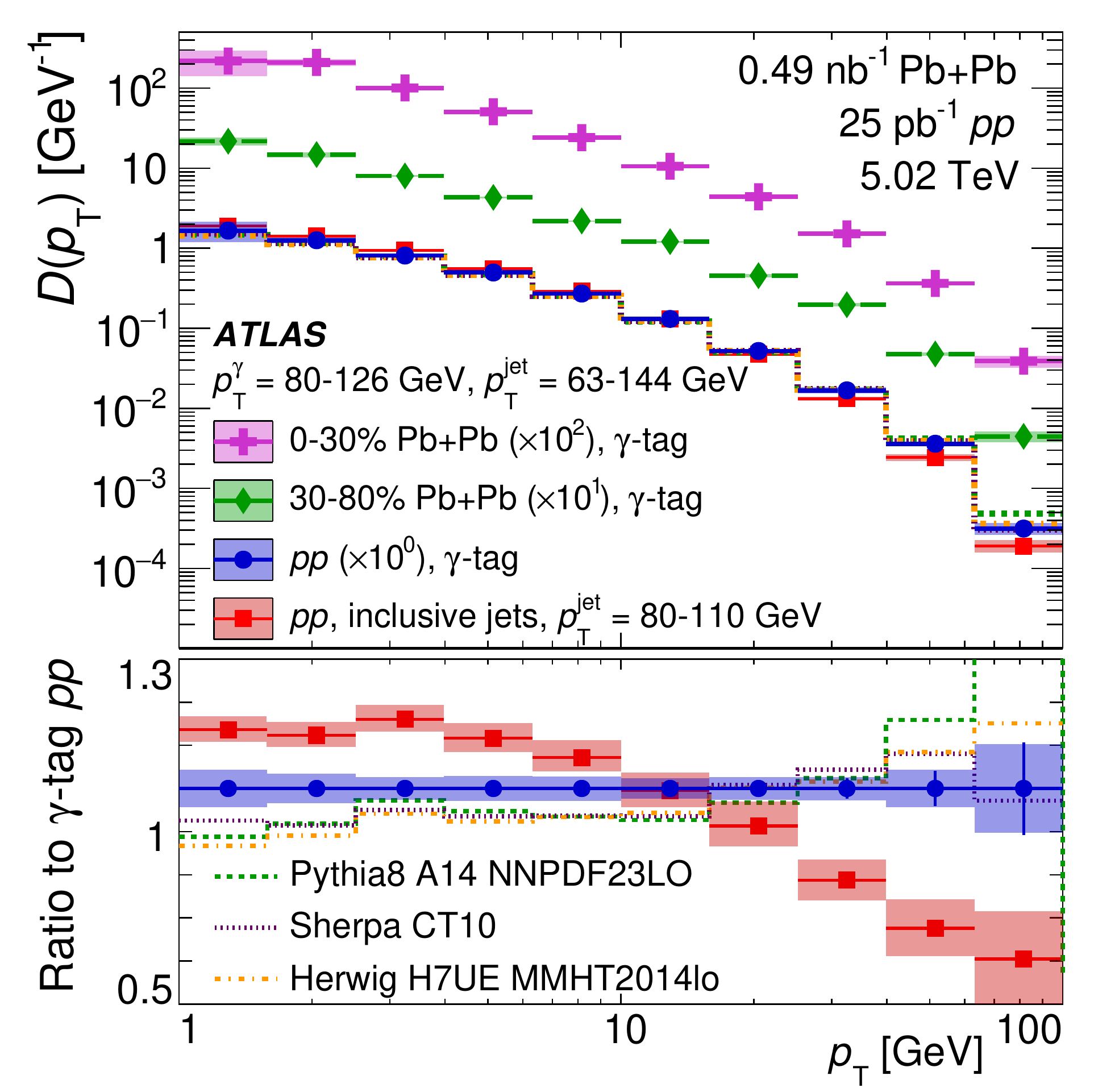}
\includegraphics[width=0.49\textwidth]{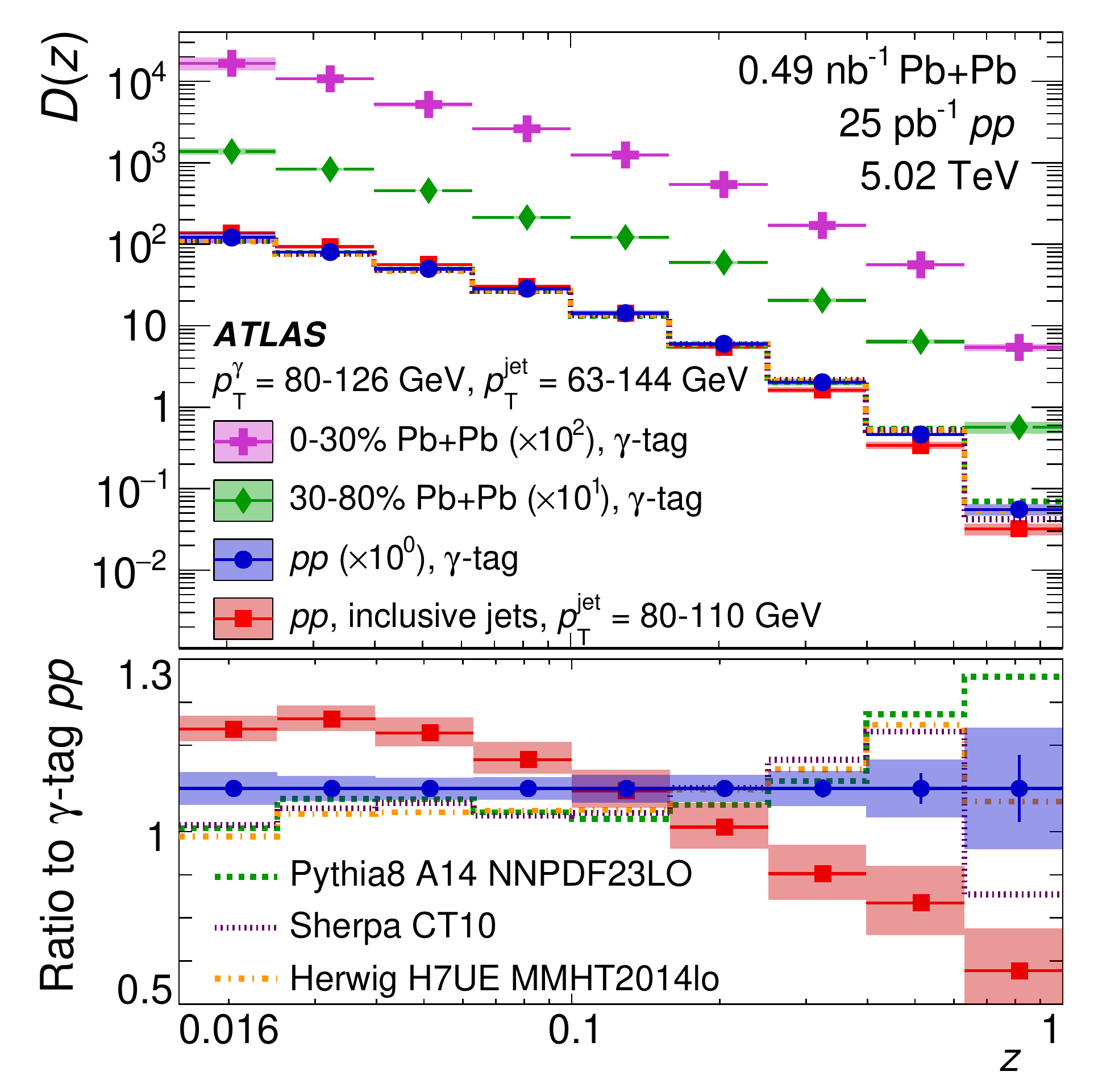}
\caption{ Fragmentation function (FF) in $\gamma$-tagged jets in $pp$ 
  events, and in central and peripheral Pb+Pb events, as a function of
  charged-particle transverse momentum \pT (left) and longitudinal
  momentum fraction $z$ (right).  The \pp results are compared with
  the analogous distribution in MC generators (dashed lines) and with
  the FF for inclusive jets in a similar \pTjet range (red
  squares). The shaded bands correspond to the total systematic
  uncertainties in the data. The bottom panels show the ratios of MC
  distributions and inclusive jet data, in \pp collisions, to the
  $\gamma$-tagged jet data, with these data plotted at unity.
\label{fig:1}}
\end{center}
\end{figure*}

The measurement and correction of the $\pT^\mathrm{jet}$ is affected
by uncertainties in the jet energy scale and resolution, which are
evaluated following the procedure~\cite{HION-2017-10} used in previous
ATLAS measurements of heavy-ion collisions.  The fake photon
background subtraction is sensitive to the determination of the photon
purity, which is evaluated as in
Ref.~\cite{HION-2018-06}. Uncertainties related to the
charged-particle yield measurement are described in detail in
Ref.~\cite{HION-2017-04}. The sensitivity to the unfolding and physics
modeling is determined through a pseudoexperiment resampling of the
response matrices, varying the prior distributions used in the
unfolding, and using the \SHERPA simulation instead of \PYTHIA8 to
perform the unfolding. For uncertainty sources with up/down
variations, the changes in the results are averaged to make a
symmetric uncertainty. For those with a single variation, an identical
uncertainty in the opposite direction is assigned.

Many of these variations change $N_\mathrm{jet}$ and $N_\mathrm{ch}$
in a significant but highly correlated way, with the result that the
FFs are less sensitive to them. Furthermore, most uncertainties are
correlated between the \pp and Pb+Pb systems, and these partially
cancel out when they are evaluated for the ratios of FFs. The total
uncertainties in the $D(z)$ and $D(\pT)$ distributions and their
ratios are typically $5$\% at moderate $z$ or \pT values. At low \pT
or $z$, the track-related uncertainties rise sharply due to the high
occupancies in Pb+Pb events. At large \pT or $z$, where the FF is very
steeply falling, the uncertainties related to the choice of prior and
physics models dominate.

Figure~\ref{fig:1} shows the corrected $D(\pT)$ and $D(z)$
distributions for jets azimuthally balanced by a high-\pT photon in
\pp events, and in central and peripheral Pb+Pb events. The 
$\gamma$-tagged jet FF in \pp collisions is observed to be harder than
the FF for inclusive jets at the same collision energy with $\pTjet$
in the range $80$--$110$~\GeV, coinciding with the peak of the
$\gamma$-tagged \pTjet distribution~\cite{HION-2015-04}. This is
consistent with the two samples having different quark jet fractions,
and with expectations from, for example, data from the Large
Electron--Positron
collider~\cite{OPAL:1995ab,Barate:1998cp,Abbiendi:1999pi}, where
harder FFs for quark jets were observed compared with those for gluon
jets.  The \pp data are also compared with generator distributions,
which are typically compatible with the data at low to moderate values
of $z$ or \pT within uncertainties.

\begin{figure*}[t]
\begin{center}
\includegraphics[width=0.99\textwidth]{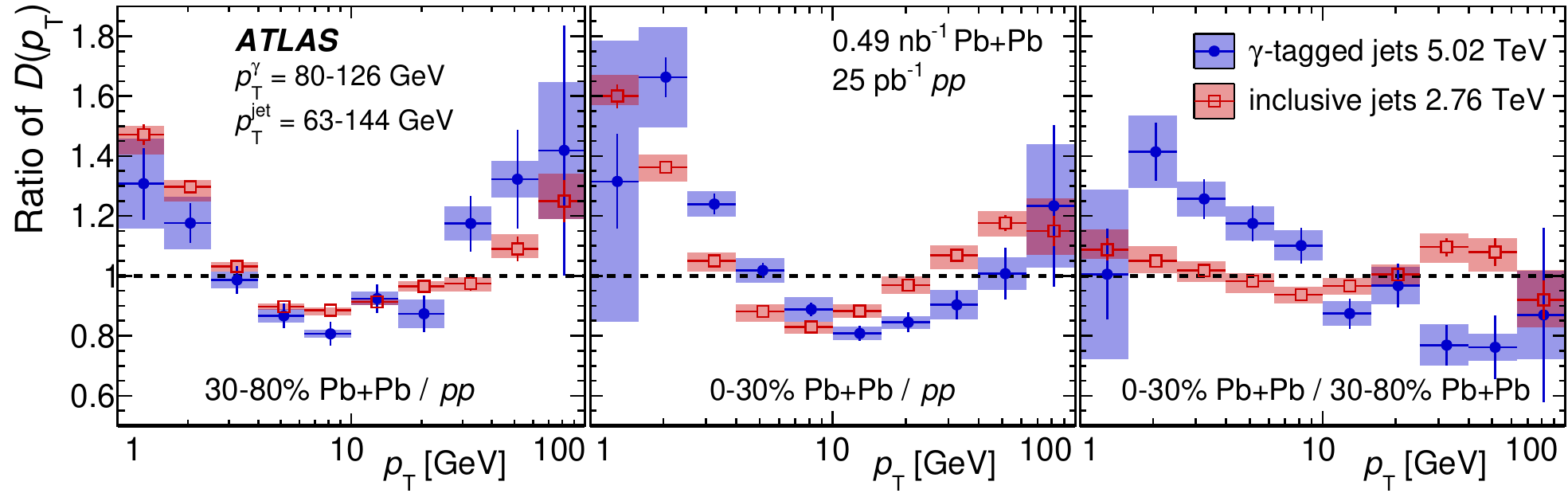}
\includegraphics[width=0.99\textwidth]{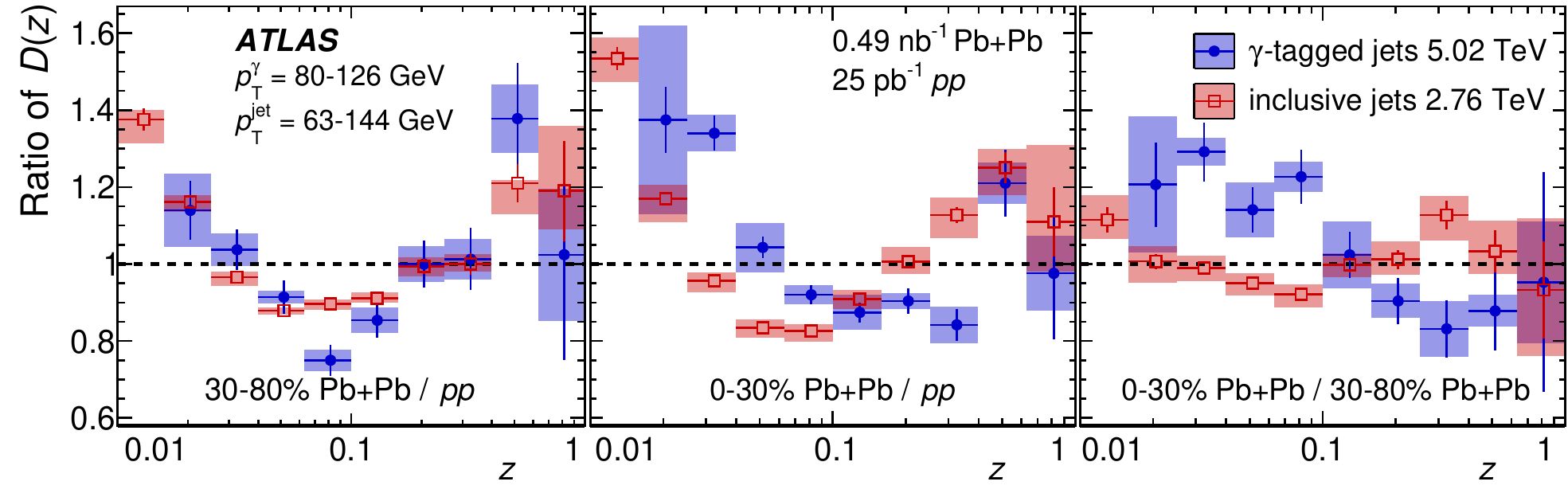}
\caption{Ratio of the fragmentation function in jets azimuthally
  balanced by a high-\pT photon: 30--80\% Pb+Pb collisions to \pp
  collisions (left panels); 0--30\% Pb+Pb collisions to \pp collisions
  (central panels); and 0--30\% to 30--80\% Pb+Pb collisions (right
  panels). Results are shown as a function of charged-particle
  transverse momentum \pT (top panels) or longitudinal momentum
  fraction $z$ (bottom panels), for $\gamma$-tagged jets (this
  measurement, full markers) and for inclusive jets in $2.76$~\TeV\
  Pb+Pb collisions~\cite{HION-2012-08,HION-2015-05} (see text, open
  markers). The centrality selections for the inclusive jet data are
  0--10\% (left), 30--40\% (central), and (0--10\%)/(30--40\%)
  (right). Hatched bands and vertical bars show for each measurement
  the total systematic and statistical uncertainties, respectively.
\label{fig:2}}
\end{center}
\end{figure*}

The left and central panels of Figure~\ref{fig:2} summarize ratios of
the $\gamma$-tagged FFs in Pb+Pb events to those in $pp$ events, and
compares them to those for inclusively selected jets with $\pTjet =
100$--$126$~\GeV\ measured in $2.76$~\TeV\ Pb+Pb and \pp
collisions~\cite{HION-2015-05}. Although the collision energy
and \pTjet range are slightly different than that for the
$\gamma$-tagged jet data, inclusive jet FFs in this region have been
observed to be compatible at the two energies and in nearby \pTjet
ranges within uncertainties~\cite{HION-2017-04}. Since the
inclusive-jet measurement uses different centrality ranges, the
centrality range corresponding to the top of that in the
$\gamma$-tagged measurement is chosen (i.e.\ 0--10\% for 0--30\% in
the $\gamma$-tagged case, and 30--40\% for 30--80\%). In peripheral
collisions, the modification pattern is quantitatively similar for
both sets of jets, featuring a depletion at moderate $z$ or \pT, and
an enhancement at very low and very high $z$ or \pT. However, in
central collisions, $\gamma$-tagged jets show an additional relative
suppression at high $z$ or \pT and a counterbalancing enhancement at
low $z$ or \pT. In addition, the minimum value of the Pb+Pb-to-\pp
ratio for $\gamma$-tagged jets is shifted to larger $z$ or \pT values.

To further explore the relative change in the FF between Pb+Pb event
classes, the ratio between central and peripheral collisions is shown
in the right panels of Figure~\ref{fig:2}. For $\gamma$-tagged jets,
the ratio is consistent with a decreasing linear function of $\log(z)$
or $\log(\pT)$, crossing unity at $z \approx 0.1$ or $\pT \approx
10$~\GeV. It is inconsistent with the analogous ratio for inclusive
jets, which is closer to unity. Thus, the data indicate that, in
central collisions, jets in $\gamma$-tagged events are modified in a
different way than inclusively selected jets.

\begin{figure*}[t]
\begin{center}
\includegraphics[width=0.49\textwidth]{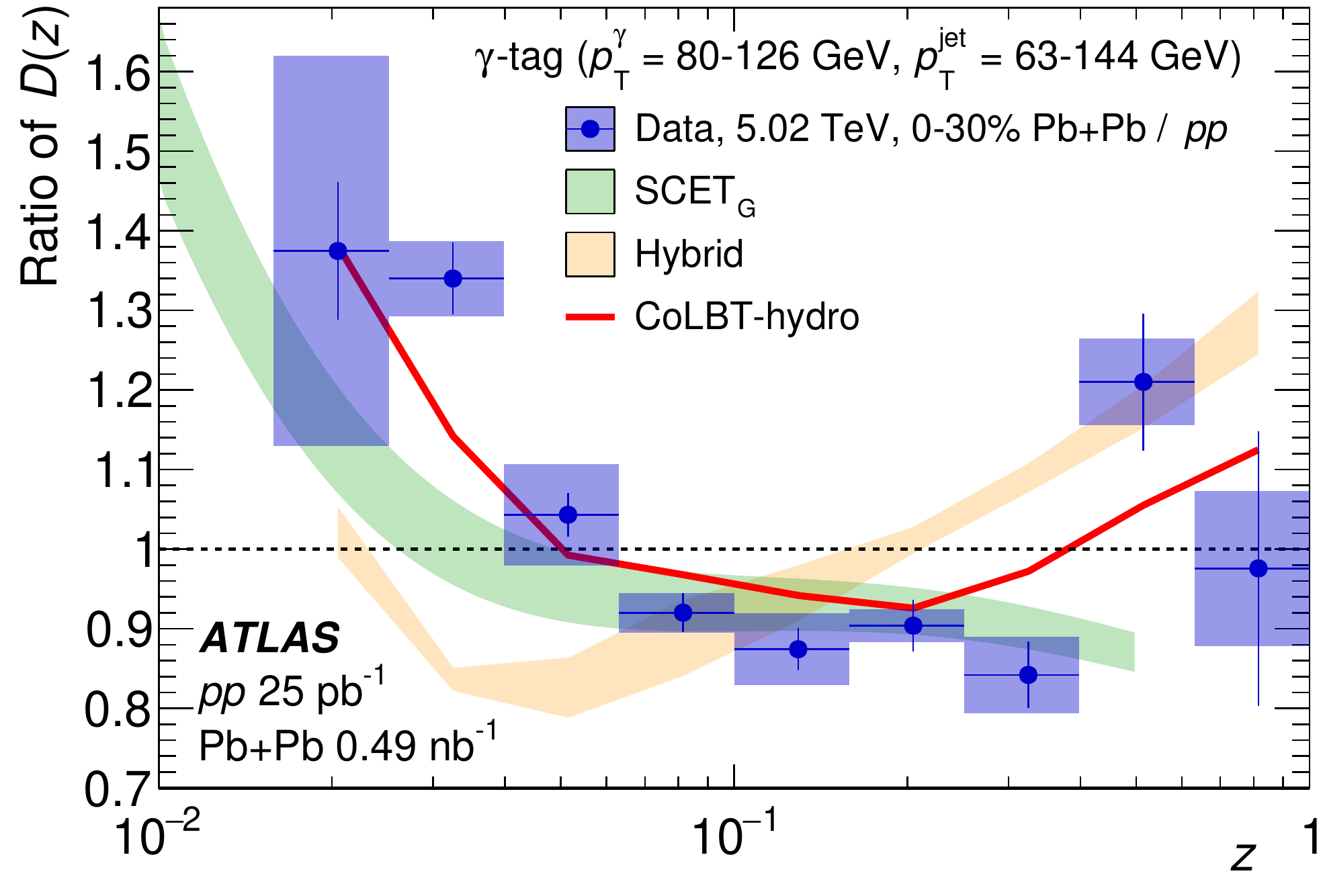}
\includegraphics[width=0.49\textwidth]{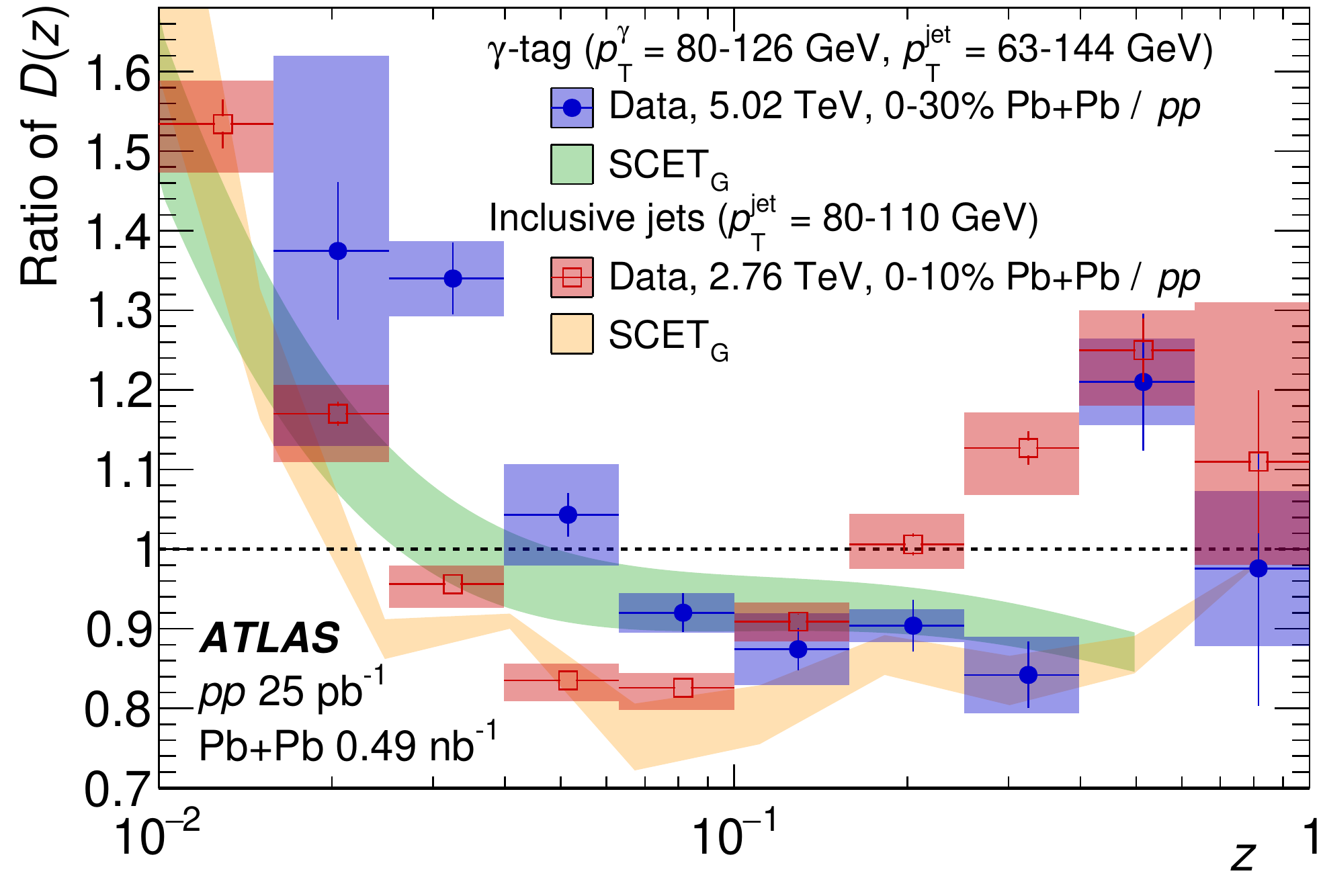}
\caption{Comparison of the ratio of $\gamma$-tagged fragmentation function
$D(z)$ in central Pb+Pb events to $pp$ events with theoretical
calculations (left). The mutual comparison between $\gamma$-tagged and
inclusive jet $D(z)$ ratios in data to each of these in the SCET$_{G}$
model is shown in the right panel. Shaded rectangles and vertical bars
show the total systematic and statistical uncertainties, respectively,
in the data.
\label{fig:3}}
\end{center}
\end{figure*}

In Figure~\ref{fig:3}, the data in central events are compared with
the results of theoretical calculations at particle level. In the left
panel, these include: (1) a perturbative calculation within the
framework of soft-collinear effective field theory with Glauber gluons
(SCET$_\mathrm{G}$) in the soft-gluon-emission (energy-loss) limit,
with jet-medium coupling $g =
2.1\pm0.1$~\cite{Kang:2017frl,Chien:2015vja}, (2) the Hybrid
Strong/Weak Coupling model~\cite{Casalderrey-Solana:2015vaa}, which
combines initial production using \PYTHIA with a parameterization of
energy loss derived from holographic methods, including back reaction
effects, and (3) the linearized Boltzmann transport (CoLBT-hydro)
model~\cite{He:2015pra} of parton propagation through quark--gluon
plasma with jet-induced medium-excitation effects. The
SCET$_\mathrm{G}$ calculation and the CoLBT-hydro model successfully
capture the key features of the $\gamma$-tagged jet FF data in the
region $z < 0.5$. In the right panel, the inclusive and
$\gamma$-tagged FF ratios in data are compared with those in
SCET$_\mathrm{G}$.  The $\gamma$-tagged FF ratio is larger than the
inclusive-jet one in the region $z < 0.1$ in both data and theory.

In summary, this Letter presents a measurement of the charged-particle
fragmentation functions for jets azimuthally balanced by a high-\pT
prompt and isolated photon. The measurement is performed using
$25$~pb$^{-1}$ of $pp$ and $0.49$~nb$^{-1}$ of Pb+Pb collision data at
$5.02$~\TeV, with the ATLAS detector at the LHC. The kinematic
selections result in events with a single leading jet, a large
fraction of which are quark jets. In $pp$ collisions, the
$\gamma$-tagged jet fragmentation functions are systematically harder
than those for inclusive jets at similar \pTjet, consistent with the
larger expected fraction of quark jets in $\gamma$-tagged events. In
30--80\% centrality Pb+Pb events, $\gamma$-tagged jets are observed to
be modified through interaction with the medium, with an overall
pattern consistent with that for inclusive jets. However, jets in
$\gamma$-tagged events are modified in 0--30\% Pb+Pb events in a
manner not observed for inclusive jets. The SCET$_G$ calculation
describes this key feature of the data. However, interpreting this
observed difference is complicated by the different jet populations in
the two cases. In Pb+Pb collisions, the inclusive jet population at
fixed \pTjet is biased towards jets which have lost the least amount
of energy. In a geometric picture, such a survivor bias selects jets
produced only near the surface of the medium. This bias is largely
avoided for $\gamma$-tagged jets, which can be selected based on the
photon kinematics. Thus they may include jets that are more quenched
on average than inclusively selected jets, including ones which sample
particularly large of path lengths.

\section*{Acknowledgments}

\input{Acknowledgements}

\clearpage


\printbibliography

\clearpage 

\input{atlas_authlist}

\end{document}

%% file: Acknowledgements.tex

We thank CERN for the very successful operation of the LHC, as well as the
support staff from our institutions without whom ATLAS could not be
operated efficiently.

We acknowledge the support of ANPCyT, Argentina; YerPhI, Armenia; ARC, Australia; BMWFW and FWF, Austria; ANAS, Azerbaijan; SSTC, Belarus; CNPq and FAPESP, Brazil; NSERC, NRC and CFI, Canada; CERN; CONICYT, Chile; CAS, MOST and NSFC, China; COLCIENCIAS, Colombia; MSMT CR, MPO CR and VSC CR, Czech Republic; DNRF and DNSRC, Denmark; IN2P3-CNRS, CEA-DRF/IRFU, France; SRNSFG, Georgia; BMBF, HGF, and MPG, Germany; GSRT, Greece; RGC, Hong Kong SAR, China; ISF and Benoziyo Center, Israel; INFN, Italy; MEXT and JSPS, Japan; CNRST, Morocco; NWO, Netherlands; RCN, Norway; MNiSW and NCN, Poland; FCT, Portugal; MNE/IFA, Romania; MES of Russia and NRC KI, Russian Federation; JINR; MESTD, Serbia; MSSR, Slovakia; ARRS and MIZ\v{S}, Slovenia; DST/NRF, South Africa; MINECO, Spain; SRC and Wallenberg Foundation, Sweden; SERI, SNSF and Cantons of Bern and Geneva, Switzerland; MOST, Taiwan; TAEK, Turkey; STFC, United Kingdom; DOE and NSF, United States of America. In addition, individual groups and members have received support from BCKDF, CANARIE, CRC and Compute Canada, Canada; COST, ERC, ERDF, Horizon 2020, and Marie Sk{\l}odowska-Curie Actions, European Union; Investissements d' Avenir Labex and Idex, ANR, France; DFG and AvH Foundation, Germany; Herakleitos, Thales and Aristeia programmes co-financed by EU-ESF and the Greek NSRF, Greece; BSF-NSF and GIF, Israel; CERCA Programme Generalitat de Catalunya, Spain; The Royal Society and Leverhulme Trust, United Kingdom. 

The crucial computing support from all WLCG partners is acknowledged gratefully, in particular from CERN, the ATLAS Tier-1 facilities at TRIUMF (Canada), NDGF (Denmark, Norway, Sweden), CC-IN2P3 (France), KIT/GridKA (Germany), INFN-CNAF (Italy), NL-T1 (Netherlands), PIC (Spain), ASGC (Taiwan), RAL (UK) and BNL (USA), the Tier-2 facilities worldwide and large non-WLCG resource providers. Major contributors of computing resources are listed in Ref.~\cite{ATL-GEN-PUB-2016-002}.

%% file: atlas_authlist.tex
 
\begin{flushleft}
{\Large The ATLAS Collaboration}

\bigskip

M.~Aaboud$^\textrm{\scriptsize 34d}$,    
G.~Aad$^\textrm{\scriptsize 99}$,    
B.~Abbott$^\textrm{\scriptsize 125}$,    
O.~Abdinov$^\textrm{\scriptsize 13,*}$,    
B.~Abeloos$^\textrm{\scriptsize 129}$,    
D.K.~Abhayasinghe$^\textrm{\scriptsize 91}$,    
S.H.~Abidi$^\textrm{\scriptsize 164}$,    
O.S.~AbouZeid$^\textrm{\scriptsize 39}$,    
N.L.~Abraham$^\textrm{\scriptsize 153}$,    
H.~Abramowicz$^\textrm{\scriptsize 158}$
% ...
\bigskip
\\

% ...
$^{13}$Institute of Physics, Azerbaijan Academy of Sciences, Baku; Azerbaijan.\\
% ...
$^{34}$$^{(a)}$Facult\'e des Sciences Ain Chock, R\'eseau Universitaire de Physique des Hautes Energies - Universit\'e Hassan II, Casablanca;$^{(b)}$Centre National de l'Energie des Sciences Techniques Nucleaires (CNESTEN), Rabat;$^{(c)}$Facult\'e des Sciences Semlalia, Universit\'e Cadi Ayyad, LPHEA-Marrakech;$^{(d)}$Facult\'e des Sciences, Universit\'e Mohamed Premier and LPTPM, Oujda;$^{(e)}$Facult\'e des sciences, Universit\'e Mohammed V, Rabat; Morocco.\\
% ...
$^{39}$Niels Bohr Institute, University of Copenhagen, Copenhagen; Denmark.\\
% ...
$^{91}$Department of Physics, Royal Holloway University of London, Egham; United Kingdom.\\
% ...
$^{99}$CPPM, Aix-Marseille Universit\'e, CNRS/IN2P3, Marseille; France.\\
% ...
$^{125}$Homer L. Dodge Department of Physics and Astronomy, University of Oklahoma, Norman OK; United States of America.\\
% ...
$^{129}$LAL, Universit\'e Paris-Sud, CNRS/IN2P3, Universit\'e Paris-Saclay, Orsay; France.\\
% ...
$^{153}$Department of Physics and Astronomy, University of Sussex, Brighton; United Kingdom.\\
% ...
$^{158}$Raymond and Beverly Sackler School of Physics and Astronomy, Tel Aviv University, Tel Aviv; Israel.\\
% ...
$^{164}$Department of Physics, University of Toronto, Toronto ON; Canada.\\
% ...
$^{a}$ Also at Borough of Manhattan Community College, City University of New York, NY; United States of America.\\
$^{b}$ Also at California State University, East Bay; United States of America.\\
$^{c}$ Also at Centre for High Performance Computing, CSIR Campus, Rosebank, Cape Town; South Africa.\\
$^{d}$ Also at CERN, Geneva; Switzerland.\\
$^{e}$ Also at CPPM, Aix-Marseille Universit\'e, CNRS/IN2P3, Marseille; France.\\
% ...
$^{*}$ Deceased

\end{flushleft}
